%% file: main.tex
\documentclass[10pt, journal]{IEEEtran}

\input{Preamble}

\input{Commands}
\input{Acronyms}

\begin{document}
\bstctlcite{CustomBSTcontrol}

\title{Toward Attention-based TinyML: \\ A Heterogeneous Accelerated Architecture and Automated Deployment Flow}
\author{
        Philip Wiese~\orcidlink{0009-0001-7214-2150},~\IEEEmembership{Graduate Student Member,~IEEE},
        Gamze \.{I}slamo\u{g}lu~\orcidlink{0000-0002-5129-1691},~\IEEEmembership{Graduate Student Member,~IEEE},
        Moritz Scherer~\orcidlink{0000-0002-2762-2307},~\IEEEmembership{Graduate Student Member,~IEEE},
        Luka Macan~\orcidlink{0009-0007-6130-8841},~\IEEEmembership{Graduate Student Member,~IEEE},
        Victor J.B. Jung~\orcidlink{0009-0001-7462-3468},~\IEEEmembership{Graduate Student Member,~IEEE},
        Alessio Burrello~\orcidlink{0000-0002-6215-8220},~\IEEEmembership{Member,~IEEE},
        Francesco Conti~\orcidlink{0000-0002-7924-933X},~\IEEEmembership{Member,~IEEE},
        Luca Benini~\orcidlink{0000-0001-8068-3806},~\IEEEmembership{Fellow,~IEEE}%
}

\maketitle

\input{src/00_Abstract}

\IEEEpeerreviewmaketitle

\input{src/10_Introduction}
\input{src/20_Background}
\input{src/30_Architecture}

\input{src/40_Implementation}
\input{src/50_Results}
\input{src/60_Conclusion}
\input{src/90_Acknowledgement}

\bibliographystyle{IEEEtran}
\bibliography{\jobname, ./references}

\input{src/95_Authors}
\vfill
\cleardoublepage

\end{document}

%% file: Preamble.tex
\usepackage{graphicx}
\usepackage[hidelinks]{hyperref}
\usepackage[free-standing-units,per-mode=symbol,mode=text]{siunitx}
\usepackage{amssymb,amsfonts}
\usepackage{amsmath}
\usepackage[export]{adjustbox}
\usepackage{float}
\usepackage{textgreek}
\usepackage{url}
\usepackage[noadjust]{cite}
\usepackage{orcidlink}
\usepackage[acronym]{glossaries}
\usepackage{tabularx}
\usepackage{enumerate}
\usepackage{threeparttable}
\usepackage{multirow}
\usepackage{booktabs} 
\usepackage{xcolor}   
\usepackage{pifont}
\usepackage{verbatim}
\usepackage{wrapfig}
\usepackage[
subtle, 
margins=tight, 
sections=normal, 
title=normal,
tracking=normal,
]{savetrees}

\DeclareSIUnit\op{Op}
\DeclareSIUnit\gateequivalent{GE}
\DeclareSIUnit\inference{Inf}

%% file: Commands.tex
\newcommand{\transformer}{Transformer}
\newcommand{\transformers}{Transformers}
\newcommand{\softmax}{Softmax}

\newcommand{\skipAuthor}{-3em}

%% file: Acronyms.tex
\newacronym{tinyML}{tinyML}{Tiny Machine Learning}
\newacronym{ita}{ITA}{Integer Transformer Accelerator}
\newacronym{e2e}{E2E}{End-to-End}
\newacronym{hwpe}{HWPE}{Hardware Processing Engine}
\newacronym{fsm}{FSM}{Finite State Machine}
\newacronym{wfi}{WFI}{Wait For Interrupt}
\newacronym{mlp}{MLP}{Multi-Layer Perceptron}
\newacronym{mha}{MHA}{multi-head Attention}
\newacronym{sha}{SHA}{single-head Attention}
\newacronym{pulp}{PULP}{Parallel Ultra Low Power}
\newacronym{axi}{AXI}{Advanced eXtensible Interface}
\newacronym{fp}{FP}{Floating Point}
\newacronym[firstplural={first-in, first-out (FIFO) buffers},longplural={first-in, first-out buffers}, first=AA]{fifo}{FIFO}{first-in, first-out}
\newacronym{DA}{DA}{Denominator Accumulation}
\newacronym{DI}{DI}{Denominator Inversion}
\newacronym{EN}{EN}{Element Normalization}
\newacronym[]{fd_soi}{FD-SOI}{fully-depleted silicon-on-insulator}
\newacronym[plural=DNNs, firstplural=Deep Neural Networks (DNNs)]{dnn}{DNN}{Deep Neural Network}
\newacronym[plural=SoCs, firstplural=Systems-on-Chip (SoCs)]{soc}{SoC}{System-on-Chip}
\newacronym[plural=ONNXs, firstplural=Open Neural Network Exchanges (ONNXs)]{onnx}{ONNX}{Open Neural Network Exchange}
\newacronym[plural=TCFs, firstplural=Tile Constraint Flows (TCFs)]{tcf}{TCF}{Tile Constraint Flow}
\newacronym[plural=TCs, firstplural=Tile Constraints (TCs)]{tc}{TC}{Tile Constraint}
\newacronym[plural=GEMMs, firstplural=General Matrix Multiplications (GEMMs)]{gemm}{GEMM}{General Matrix Multiplication}
\newacronym{cp}{CP}{Constraint Program}
\newacronym[plural=QNNs, firstplural=Quantized Neural Networks (QNNs)]{qnn}{QNN}{Quantized Neural Network}
\newacronym[plural=FMs, firstplural=Foundation Models (FMs)]{fm}{FM}{Foundation Model}
\newacronym{xr}{XR}{Extended Reality}
\newacronym[plural=RNNs, firstplural=Recurrent Neural Networks (RNNs)]{rnn}{RNN}{Recurrent Neural Network}
\newacronym[plural=CNNs, firstplural=Convolutional Neural Networks (CNNs)]{cnn}{CNN}{Convolutional Neural Network}
\newacronym{nlp}{NLP}{Natural Language Processing}
\newacronym{hpc}{HPC}{High-Performance Computing}
\newacronym{eeg}{EEG}{Electroencephalography}
\newacronym[plural=LLMs, firstplural=Large Language Models (LLMs)]{llm}{LLM}{Large Language Model}
\newacronym[plural=SLMs, firstplural=Small Language Models (SLMs)]{slm}{SLM}{Small Language Model}
\newacronym[plural=GPGPUs, firstplural=General-Purpose Graphics Processing Units (GPGPUs)]{gpgpu}{GPGPU}{General-Purpose Graphics Processing Unit}
\newacronym[plural=TPUs, firstplural=Tensor Processing Units (TPUs)]{tpu}{TPU}{Tensor Processing Unit}
\newacronym[plural=DMAs, firstplural=Direct Memory Access (DMAs)]{dma}{DMA}{Direct Memory Access}
\newacronym{ai}{AI}{Artificial Intelligence}
\newacronym[plural=MLs, firstplural=Machine Learnings (MLs)]{ml}{ML}{Machine Learning}
\newacronym[plural=MMUs, firstplural=Memory-Management Units (MMUs)]{mmu}{MMU}{Memory-Management Unit}
\newacronym[plural=ISAs, firstplural=Instruction Set Architectures (ISAs)]{isa}{ISA}{Instruction Set Architecture}
\newacronym{simd}{SIMD}{Single Instruction Multiple Data}
\newacronym{spmd}{SPMD}{Single Program Multiple Data}
\newacronym{dsp}{DSP}{Digital Signal Processing}
\newacronym[plural=MCUs, firstplural=Microcontrollers (MCUs)]{mcu}{MCU}{Microcontroller}
\newacronym{tcdm}{TCDM}{Tightly-Coupled Data Memory}
\newacronym{nms}{NMS}{Neural Memory Subsystem}
\newacronym{npu}{NPU}{Neural Processing Unit}
\newacronym{mram}{MRAM}{Magnetoresistive Random Access Memory}
\newacronym{sram}{SRAM}{Static Random Access Memory}
\newacronym{ptq}{PTQ}{Post-Training Quantization}
\newacronym{qat}{QAT}{Quantization-Aware Training}
\newacronym{efm}{EFM}{Embodied Foundation Model}
\newacronym{ast}{AST}{Abstract Syntax Tree}
\newacronym{tqt}{TQT}{Trained Quantization Thresholds}
\newacronym{api}{API}{Application Programming Interface}
\newacronym{cim}{CIM}{Compute-In Memory}
\newacronym{mps}{MPS}{Metal Performance Shaders}
\newacronym[plural=MPUs, firstplural=Microprocessors (MPUs)]{mpu}{MPU}{Microprocessor}
\newacronym{dram}{DRAM}{Dynamic Random Access Memory}
\newacronym{ir}{IR}{Intermediate Representation}\newacronym{cv}{CV}{Computer Vision}
\newacronym[longplural={Vision Transformers}]{ViT}{ViT}{Vision Transformer}
\newacronym{hwce}{HWCE}{Hardware Convolution Engine}

%% file: src/00_Abstract.tex
\begin{abstract}
One of the challenges for \gls{tinyML} is keeping up with the evolution of \acrlong{ml} models from \acrlongpl{cnn} to Transformers.
We address this by leveraging a heterogeneous architectural template coupling RISC-V processors with hardwired accelerators supported by an automated deployment flow.
We demonstrate Attention-based models in a \gls{tinyML} power envelope with an octa-core cluster coupled with an accelerator for quantized Attention. 
Our deployment flow enables end-to-end 8-bit Transformer inference, achieving leading-edge energy efficiency and throughput of \qty[detect-all=true]{2960}{\giga\op\per\joule} and \qty[detect-all=true]{154}{\giga\op\per\second} (\qty[detect-all=true]{0.65}{\volt}, \qty[detect-all=true]{22}{\nano\meter} FD-SOI technology).
\end{abstract}

\glsresetall

%% file: src/10_Introduction.tex
\section{Introduction}
\label{sec:intro}

In recent years, \gls{tinyML} has attracted much attention, bringing compute-intensive \gls{ai} models towards deployment on \gls{mcu} class devices with power envelopes of a few milliwatts.
Embedding \glspl{dnn} in small, low-power devices is highly relevant for numerous applications ranging from multi-modal sensing and keyword spotting to anomaly detection and smart wake-up~\cite{abadade_comprehensive_2023}.
Compared with cloud-only inference, tinyML offers lower network utilization, higher privacy, and more predictable latency.
However, extreme-edge devices typically run with a tightly constrained memory budget, without fully-fledged operating systems and advanced hardware features such as \glspl{mmu} and fully automated cache hierarchies.

One of the key research challenges is whether it is possible to build systems that respect the tight hardware and software cost and power constraints of \gls{tinyML} systems while supporting the rapid advancement of models.
A key consideration for addressing this research question is the trade-off between specialization and generality on the computer architecture level. 
Although numerous model-specific accelerators have been proposed in recent years~\cite{abadade_comprehensive_2023}, designing \glspl{soc} that can integrate these accelerators while remaining adaptable to evolving \gls{ai} models remains an open challenge, particularly under tight memory constraints.
Moreover, automatically and efficiently deploying rapidly evolving \gls{dnn} models, especially the increasingly popular Attention-based networks, on accelerator-enhanced \glspl{mcu} remains a significant challenge. 
Additionally, fast product cycles make it difficult to accommodate the time and cost associated with hand-tuning each model for deployment.

In this paper, we address what we believe to be a fundamental question for the future of \gls{tinyML}:
How can we move from classical perceptive \gls{ai} and \gls{cnn} models toward leading Attention-based \transformer{} models?
Unlike in \glspl{cnn}, complex dataflow operations like \softmax{} in \transformers{} can lead to high latency despite their low arithmetic complexity. 
While \gls{gemm} accelerators handle most computations in \transformer{} networks efficiently, the remaining operations can become a bottleneck.
To address this challenge, we leverage a flexible \gls{mcu}-class architectural template for efficiently integrating specialized hardware accelerators with multi-core clusters over a low-latency \gls{tcdm} interconnect.
At its core, we use a RISC-V (RV32) compute cluster based on the latency-tolerant Snitch core~\cite{zaruba_snitch_2021}.
To the best of our knowledge, this is the first heterogeneous Snitch-based cluster integrating \glspl{hwpe}\footnote{\href{https://hwpe-doc.readthedocs.io/en/latest/index.html}{https://hwpe-doc.readthedocs.io/en/latest/index.html}}, advancing beyond previous configurations which focused on instruction extension units tightly coupled to the pipeline.
We introduce an extensible deployment flow based on a bottom-up \gls{dnn} compiler, Deeploy, that enables fast and automated \gls{e2e} deployment.
Using this template, we integrate an extended version of the \gls{ita}~\cite{islamoglu_ita_2023} and prove our hardware-software co-design flow on Attention-based models.
As a concrete use case, we showcase the \gls{e2e} deployment of MobileBERT~\cite{sun_mobilebert_2020}, DINOv2~\cite{oquab_dinov2_2023}, and Whisper's encoder~\cite{radford_robust_2023} within a power envelope of \SI{52.0}{\mW} (GlobalFoundries \SI{22}{\nm} fully-depleted silicon-on-insulator technology at \SI{0.65}{\volt}).

The contributions of this paper are as follows:
\begin{itemize}
    \item We propose a novel, flexible hardware-software architecture template designed to meet the dataflow and compute requirements of emerging Attention-based \gls{ai} workloads. Our hardware architecture allows the co-integration of a multi-core latency-tolerant Snitch compute cluster with complex hardware accelerators over a high-bandwidth, low-latency \gls{tcdm} interconnect. At the same time, our co-optimized software template facilitates efficient \gls{e2e} workload mapping.
    We demonstrate that our hardware-software template enables starvation-free contention for resources in the shared memory with its tunable interconnect bandwidth and the \gls{dma} engine.
    As a result, we achieve accelerator utilization of up to \SI{85.1}{\percent}.
    \item As a concrete use-case, we integrate \gls{ita}, a \transformer{} accelerator tuned for the specific dataflow of the Attention calculation, into our hardware-software template and extend Deeploy\footnote{\url{https://github.com/pulp-platform/Deeploy}} with an accelerator model to enable automated mapping, scheduling, tiling, and code generation. We evaluate the performance through post-layout power analysis, achieving a peak performance of \SI{741}{\giga\op\per\second} and energy efficiency of up to \SI{6.35}{\tera\op\per\joule}.
    The integration incurs only a \SI{4.7}{p{.}p{.}} decrease in utilization compared to the standalone accelerator, demonstrating the low overhead of our template.
    \item We showcase the capability of our hardware-software template to support a range of Attention-based \gls{tinyML} models, including MobileBERT, DINOv2, and Whisper’s encoder.
    Our flow unlocks the potential for collaborative execution between the cluster and the hardware accelerator, which optimizes performance and energy efficiency and prevents resource starvation. By enabling this collaborative execution, we significantly enhance \gls{e2e} inference energy efficiency by 102$\times$ compared to inference without the accelerator, achieving an \gls{e2e} throughput of up to \SI{154}{\giga\op\per\second} and energy efficiency of \SI{2.96}{\tera\op\per\joule}. 
\end{itemize}

%% file: src/20_Background.tex
\section{The Challenges of TinyML Acceleration}
\label{sec:background}

\subsection{HW Integration Challenges for Attention-based Networks}
\label{sec:background_accelerator}

Over the years, several approaches to integrating hardware accelerators into \glspl{soc} were proposed, varying in their degree of coupling to the \gls{soc}'s processor.

A well-developed approach relies on closely coupling the accelerator with the processor cores through instruction-set extensions~\cite{cui_risc-v_2023}. While this approach enables ample flexibility in workload mapping, it is inadequate for Attention accelerators that require large bandwidth. In fact, instruction extensions are limited by the core's load/store interface, the bandwidth and size of the register file, and instruction fetch bandwidth.

On the other end of the spectrum is the loosely coupled integration of accelerators with internal private memory~\cite{cota_analysis_2015}.
While this approach eliminates memory access contention during inference, it requires a large in-accelerator and fully private integrated memory to store the intermediate tensors generated for Attention. This causes large area requirements, which increase the cost of the accelerator.
It also hinders collaboration between different engines, as data must be moved explicitly between memory hierarchy levels with a significant energy overhead.
An interesting middle ground between these two extremes is to couple the accelerator and cores through shared memory~\cite{cota_analysis_2015}. 
Unlike private memory solutions, this approach facilitates data exchange between the accelerators and cores. This is a key feature for Attention-based networks since it allows cores to perform auxiliary operations easily without memory copy overheads. These operations vary significantly across different model variants, often preventing hardware acceleration.

In this work, we propose a novel architectural template, integrating a cluster of RISC-V cores with an accelerator over shared L1 memory. 
We show that our proposed design enables close interaction between the cluster cores and the accelerator, supporting emerging and evolving variations of non-linearities and normalization layers found in Attention-based models while exploiting the accelerator for supported operators.

\subsection{TinyML Software Deployment Challenges}
\label{sec:background_deployment}

Deploying \transformers{} at the extreme edge on devices with hardware accelerators comes with many difficulties as they require significant software effort to unlock the performance and efficiency of the accelerators. First and foremost, \gls{tinyML} devices have highly constrained on-chip memory, in the order of \si{\mebi\byte}, and no operating systems.
Hence, one must \textit{tile} layers to process tensors from the lowest level of the memory hierarchy. 
Moreover, these systems often feature software-managed scratchpad memory hierarchies. Thus, explicit and uncached \gls{dma} transfers are required to transfer tiled tensors. Furthermore, static memory allocation is crucial to guarantee conflict-free memory transfers.

While several code generation tools for \glspl{cnn} have been demonstrated~\cite{abadade_comprehensive_2023}, most do not generalize to Attention-based models. While \glspl{cnn} use few branches in their dataflow graphs and therefore do not require sophisticated memory allocation strategies, the highly parallel and branching structure of Attention-based networks requires novel lifetime analysis and tiling strategies to effectively tile and schedule their execution. 

%% file: src/30_Architecture.tex
\section{Architecture Template}
\label{sec:architecture}

\begin{figure*}[!ht]
\input{fig/01_figure_flow}
\label{fig:cluster}
\end{figure*}

In this Section, we describe a flexible architecture template, shown in \autoref{fig:cluster}, that combines multiple \gls{dsp} optimized RISC-V cores into a compute cluster and facilitates the integration of newly developed hardware accelerators using the \gls{hwpe} infrastructure and automated deployment.
Compared to a single-core system, this enables efficient operation through higher performance and parallelism and enhances adaptability and scalability.
The \gls{hwpe} interface developed for the \gls{pulp} platform facilitates the integration of accelerators with multi-core compute cluster into a shared memory cluster.
Our template integrates the area-efficient Snitch cores, occupying \SI{22}{\kilo\gateequivalent} each~\cite{zaruba_snitch_2021}.
Snitch is a single-stage, in-order core implementing integer base RV32I, RV32M subset for integer multiply/divide instructions, and standard atomic instruction extension RV32A. Unlike CV32E40P cores used in other PULP-derived clusters\footnote{\url{https://docs.openhwgroup.org/projects/cv32e40p-user-manual}}, Snitch cores are significantly smaller (-56\%) and provide a decoupled memory interface, allowing latency-tolerant memory access by pipelining multiple loads and stores.

We couple the cores and accelerators through the shared interleaved L1 \gls{tcdm} to facilitate energy-efficient data exchange between the compute elements. This is especially crucial for rapidly evolving Attention-based networks as various auxiliary operations need to be computed on the cluster while the majority of the computation is conducted on the accelerator.
To reduce banking conflicts and provide the high bandwidth Attention accelerators need, we use 32 banks with \SI{4}{\kibi\byte} each, resulting in a total capacity of \SI{128}{\kibi\byte}.
The multi-banked memory makes it unnecessary to attach additional private memory to the accelerator, as data can be accessed by both the accelerator and the cluster's cores simultaneously. 
We use a 64-bit \gls{tcdm} interconnect, which is implemented as a combinatorial crossbar, resulting in single-cycle latency in the absence of conflicts with \SI{256}{\byte\per cycle} bandwidth towards the L1.
Each core has one master port with decoupled request and response path connected to the \gls{tcdm} interconnect, and the \gls{hwpe} subsystem features a parametric number $N_{\textrm{HWPE}}$ of master ports to allow the integration accelerators.

The cluster includes two parametrizable \gls{axi} interconnects: a wide crossbar with a $D_{\textrm{AXI,W}}$ bit data width and a narrow crossbar with a $D_{\mathrm{AXI,N}}$ bit data width.
The wide \gls{axi} interconnect is used to load instructions into the shared \SI{8}{\kibi\byte} instruction cache and to transfer data from and to the \gls{soc} level memory system in conjunction with the \gls{dma}.
The narrow \gls{axi} interconnect is intended to connect to the \gls{soc} interconnect to attach peripherals and communicate with a host system.
Moreover, one Snitch core is coupled with a \gls{dma} to manage data movements within the cluster, facilitating double buffering to maintain high accelerator utilization.

\subsection{HWPE Subsystem}\label{sec:architecture_hwpe}
The \gls{hwpe} template provides three modules: a \textit{controller}, one or multiple \textit{streamers}, and the \textit{engine}. 
The \textit{controller} is the interface between the cores in the cluster and the accelerator. 
It has a \gls{fsm} specific to the engine to govern the operation of the accelerator and a memory-mapped register file to keep parameters for the accelerator. 
The register file can hold a sequence of multiple \textit{tasks} that can be programmed by any core in the cluster through the controller interface over the narrow \gls{axi} interconnect. A \textit{task} represents a set of configuration values used by the accelerator.
The \textit{streamers} act as a special-purpose low-cost \gls{dma} to load and store data from the shared \gls{tcdm}. 
Finally, the \textit{engine} contains a hardware accelerator that accepts the streamer's data and the controller's configuration.

\gls{hwpe} allows connecting accelerators seamlessly to \gls{pulp} clusters and makes programming straightforward over the peripheral interface accessible via \gls{axi}. 
Three steps are necessary to integrate an accelerator into the \gls{hwpe} subsystem. First, the required number of streamers must be instantiated in accordance with the accelerator's data ports. Next, the streamers must be connected with the accelerator's data ports and the \gls{tcdm} interconnect. Finally, an \gls{fsm} controlling the accelerator and streamers must be implemented.

\gls{hwpe} provides two types of streamers: one for input, \textit{source streamers} and one for output, \textit{sink streamers}.
The streamers utilize a simple valid-ready handshake protocol on the accelerator side, ensuring compatibility with most accelerators. 
Additionally, \gls{hwpe} includes \glspl{fifo} on both the \gls{tcdm} and accelerator sides, which can be instantiated and sized according to the specific needs of the accelerator and cluster.
We time-multiplex multiple streamers to a multi-port interface with $N_{\textrm{HWPE}}$ ports and connect to the \gls{tcdm} interconnect.

The final step of integrating an accelerator into the \gls{hwpe} involves designing an \gls{fsm} to control both the accelerator and the streamers. 
We use a \textit{controller} that supports a programmable multi-context register file, allowing the cores to offload the next \textit{task} while the accelerator runs, thereby hiding configuration latency. 
The \gls{fsm} designed around the control slave is straightforward: it reads the configuration for the accelerator from the register file, transfers it to the engine, and configures the streamers accordingly.

\subsection{Neural Network Deployment Framework}\label{sec:deploymentframework}

To execute \transformer{} models on the proposed architectural template, we integrate our hardware template in the Deeploy compiler~\cite{scherer_deeploy_2024}, which maps neural networks to user-defined, platform-specific C code kernel templates.
Deeploy is a \gls{dnn} compiler that offers architecture-agnostic \gls{tinyML} optimizations like double-buffering, memory-aware operator tiling, \gls{dma}-aware code generation, and fully static offline memory layout generation.
These features allow us to accommodate the custom tiling required for operators exclusively present in Transformer networks. 

In this way, Deeploy generates code to offload supported \gls{dnn} operators onto accelerators while providing highly optimized fallback kernel implementations for unsupported operators on the cluster. This bottom-up approach guarantees that emerging \glspl{dnn} operators can be mapped to our general-purpose cores while fully leveraging integrated accelerators for their supported operators. This is especially useful when considering the numerous variants of Attention-based models, which contain the same Attention mechanism but have slightly different activation or normalization functions. 

To integrate a new \gls{hwpe} accelerator, Deeploy only requires a minimal accelerator model; first, the accelerator model must specify the geometrical tiling constraints for operators it can run. Second, the model must provide minimal arithmetic templates for running each supported operator. All other necessary performance optimizations, including memory-aware operator tiling, static memory layout generation, double-buffering code generation, and \gls{dma}-aware memory transfers, are inserted by Deeploy automatically. 

By integrating a model of the hardware template with Deeploy, we propose a low-overhead, adaptable hardware-software architecture template that minimizes the development effort for both hardware and software integration while meeting the strict requirements of extreme edge Attention-based model deployment.

%% file: fig/01_figure_flow.tex
\centering
\includegraphics[width=0.95\linewidth]{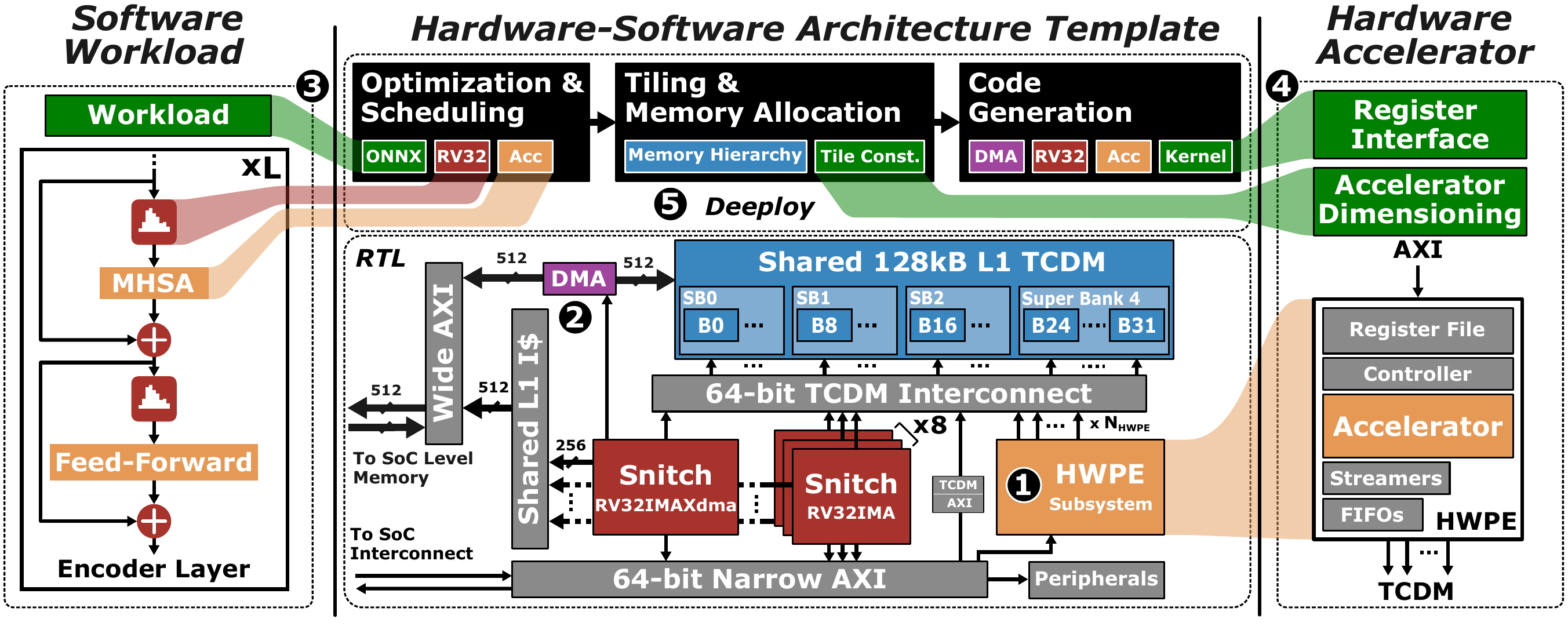}
\caption{Overview of the Hardware-Software Architecture Template. 
The flexible template allows modular integration of accelerators into an \gls{soc} and deployment of different workloads with Deeploy.
The workflow is as follows: \includegraphics[height=1em]{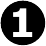} Integrate an accelerator as an \gls{hwpe} engine, a configurable interface designed for efficient integration of memory-coupled accelerators, enabling streamlined data transfer and control between the accelerator and shared memory. \includegraphics[height=1em]{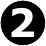} Ensure sufficient bandwidth for the accelerator by tuning the wide \gls{axi} interconnect, allowing high-bandwidth access to L2 memory via the \gls{dma}. 
\includegraphics[height=1em]{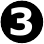} Configure the operator mapping in Deepooy and provide the workload as an \gls{onnx} graph. \includegraphics[height=1em]{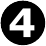} Define the tiling constraints according to the accelerator buffer and datapath sizes and provide minimal kernel templates to control the accelerator via a register interface. \includegraphics[height=1em]{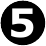} Use Deeploy to perform automated graph optimization and scheduling, to co-optimize operator tiling and static memory allocation, and to generate C code. This code orchestrates memory transfers using the \gls{dma} and coordinates execution on the compute cores and the accelerator.}

%% file: src/40_Implementation.tex
\section{Implementation}
\label{sec:implementation}

\begin{figure*}[!ht]
    \centering
    \includegraphics[width=0.95\linewidth]{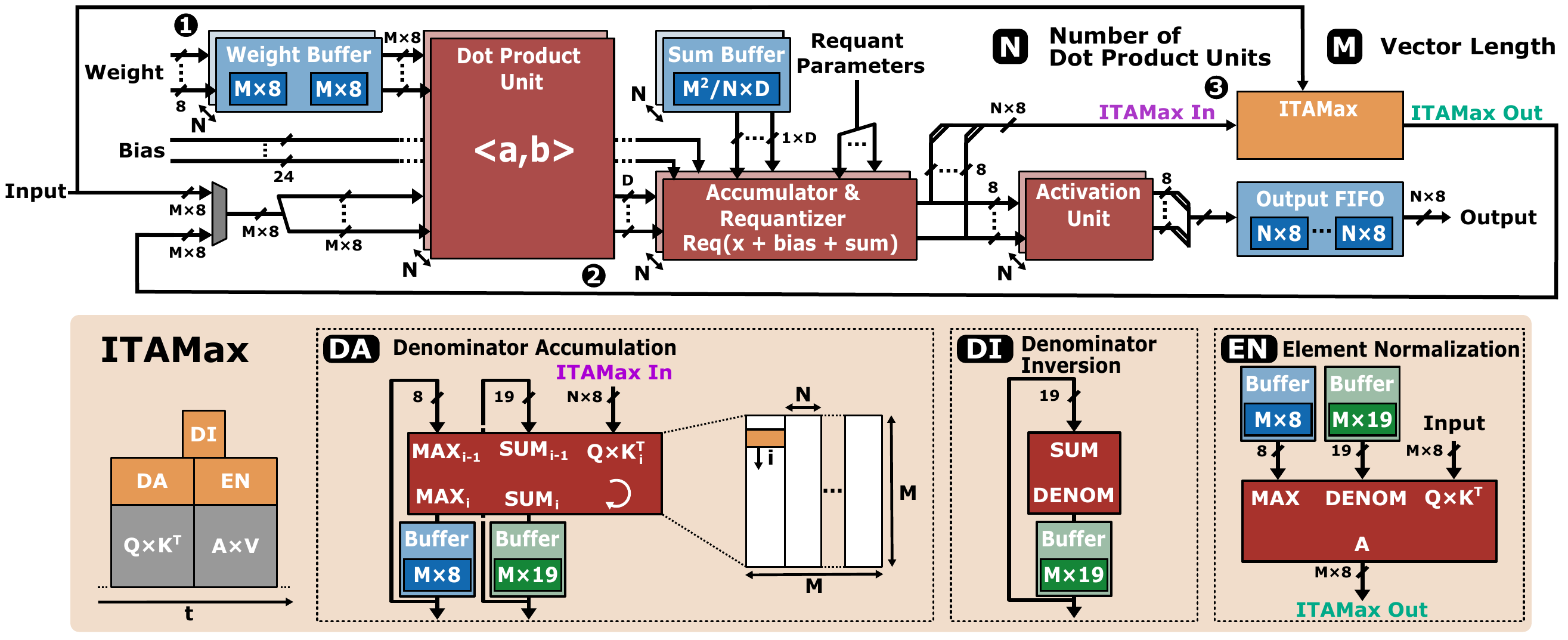}
    \caption{Architecture of the Integer Transformer Accelerator (ITA). ITA combines an output stationary dataflow with a local weight stationary dataflow and streaming \softmax{} operation to achieve high data reuse and minimal memory interaction. \includegraphics[height=1em]{fig/icons/one_icon.pdf} Weights are stored in a double-buffered weight memory to fetch the next set of weights while performing computation with the current set of weights. \includegraphics[height=1em]{fig/icons/two_icon.pdf} Inputs are fetched via streamers and passed through the ITAMax module during $\mathbf{A \times V}$ step. \includegraphics[height=1em]{fig/icons/three_icon.pdf} While $\mathbf{Q} \times \mathbf{K}^\mathrm{T}$ is computed, the ITAMax module operates on the outputs to accumulate the denominator. ITAMax operates in three stages: \includegraphics[height=1em]{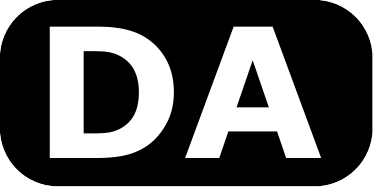} Find the local maximum and compare it with the previous maximum stored in the buffer, accumulate the denominator of the \softmax{} using the current maximum and normalize the previous sum if the maximum is changed. \includegraphics[height=1em]{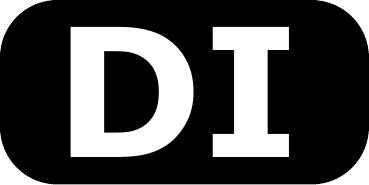} After the accumulation, the denominator is inverted and saved to the same buffer. \includegraphics[height=1em]{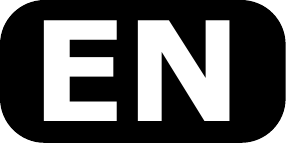} Inputs for $\mathbf{A \times V}$ step are normalized using the saved maximum and inverted denominator.}
    \label{fig:ita}
\end{figure*}

As a concrete implementation of our template, we show a platform that couples a cluster with 8+1 RV32IMA Snitch cores with the \acrfull{ita}~\cite{islamoglu_ita_2023}.
The \gls{ita} accelerator enables the acceleration of 8-bit \gls{gemm} and the more complex \gls{mha} present in \transformer{} networks. \gls{ita} used in this work is an extended version of the accelerator presented in \cite{islamoglu_ita_2023}, featuring additional functionality through the inclusion of a partial sum buffer and an activation unit supporting ReLU and GeLU. Furthermore, it is wrapped with \gls{hwpe} components.

\subsection{Integer Transformer Accelerator (ITA)}
\gls{ita} is an accelerator for encoder-only \transformer{} models and performs efficient inference in 8-bit arithmetic, using an integer-only \softmax{} approximation.
\autoref{fig:ita} shows the architecture of \gls{ita}. 
At the core of \gls{ita}, there are $N$ dot product units that compute the dot product between two vectors of length $M$. 

\Gls{ita} integrates a \softmax{} approximation, referred to as \textit{ITAMax}, that operates on integer values in a streaming mode.
This enables computing \softmax{} on the fly.
\softmax{} is defined as
\begin{equation}
    \text{Softmax}(\boldsymbol{x})_i = \frac{e^{x_i-\max(\boldsymbol{x})}}{\sum_{j=1}^{n} e^{x_j-\max(\boldsymbol{x})}}
\end{equation}
and normalizes the input matrix row-wise, transforming them into probabilities.
This is used in \transformers{} to calculate the Attention $\mathbf{A \times V}$ with
\begin{equation}
    \mathbf{A \times V} = \mathrm{Softmax}\left(\mathbf{Q} \times \mathbf{K}^\mathrm{T}\right) \times \mathbf{V}
\end{equation}\label{eq:attention}

The \textit{ITAMax} unit has three stages of operation as illustrated in \autoref{fig:ita}. 
The first \gls{DA} stage operates on the 8-bit quantized dot product results from the $\mathbf{Q} \times \mathbf{K}^\mathrm{T}$ matrix multiplication. 
It determines the maximum of the partial row results and accumulates the denominator of the \softmax{} with the current maximum. The current maximum and the accumulated denominator are stored in buffers.
At every iteration, if the local row maximum differs from the previous one, the partial sum is renormalized, and the global maximum is updated.

Once \textit{ITAMax} processes the entire row and accumulates the denominator with the global maximum of the row, it inverts the denominator in the \gls{DI} stage and stores it internally. 
The \gls{EN} stage only starts when the post-\softmax{} activations are required as input to \gls{ita} in the next matrix multiplication ($\mathbf{A} \times \mathbf{V}$).
This stage normalizes the values from the $\mathbf{Q} \times \mathbf{K}^{\mathrm{T}}$ calculation on the fly to produce $\mathbf{A}$.

With this unique dataflow, \textit{ITAMax} performs \softmax{} without additional latency and data fetching from the L1 memory with a low area and power overhead.
Since \gls{ita} integrates a datapath for single-head Attention, \gls{mha} must be calculated sequentially head-by-head. Therefore, \gls{ita} operates on a single head at a time and computes the partial output projection for each head. The partial outputs of each head need to be summed by the external cluster.

Additionally, \gls{ita} integrates activation units that fully operate in integer arithmetic. 
The activation unit has three modes of operation: Identity, ReLU, and GeLU, which can be selected for each computation via the configuration interface of \gls{hwpe}. 
For the integer approximation of GeLU, we use the i-GeLU~\cite{kim_i-bert_2021} performed in $D$-bit and quantized the results to 8-bit.
This allows using \gls{ita} as a \gls{gemm} accelerator with activation functions accelerated in hardware.

\subsection{Accelerator Integration}
\label{sec:implementation_ita}

For \gls{ita}, we use $N=16$ dot product units with a $D=26$-bit accumulator to support matrix dimensions up to 512 and a vector length of $M=64$.
We choose this configuration to exploit the memory-side bandwidth the \gls{tcdm} offers.
As \gls{ita} features three input ports (input, weight, bias) and one output port, three input streamers and one output streamer are required. 

As the four streamers are multiplexed in time, \gls{ita} requires \SI{128}{\byte\per cycles} of maximum bandwidth to fetch two input vectors per cycle; therefore, we use 16 master ports on the \gls{tcdm} interconnect for the \gls{hwpe} subsystem.
To produce one output tile, \gls{ita} takes at least \SI{256}{cycles} and the \gls{dma} needs to fetch at most two $64 \times 64$ 8-bit inputs/weights, $64$ 24-bit bias values and write back $64 \times 64$ 8-bit outputs from and to the L2 memory.
This results in a worst-case average bandwidth of \SI{48.75}{\byte\per cycles} towards the \gls{soc} memory.
Consequently, we use a 512-bit wide data \gls{axi} interconnect to provide enough bandwidth for the instructions cache and \gls{ita}.
Moreover, we use 64-bit for the narrow \gls{axi} interconnect to enable the integration of the cluster into a 64-bit host system.
Finally, in \gls{ita}, we use a dual-context register file that can be programmed via the narrow 64-bit \gls{axi} interconnect.
As the \gls{hwpe} \textit{Controller} uses the peripheral interface, we place an adapter between the \gls{axi} bus and the module.

\subsection{Physical Implementation}\label{sec:implementation_physical}

To evaluate our architecture in a \gls{tinyML}-friendly technology node, we implemented the complete Snitch cluster with an extended version of the  \gls{ita} accelerator in GlobalFoundries' \SI{22}{\nano\meter}\,FDX \gls{fd_soi} technology, targeting an operating frequency of \SI{500}{\mega\hertz} under typical conditions (TT, \SI{0.8}{\volt}, \SI{25}{\celsius}), and \SI{425}{\mega\hertz} in the energy-efficient core voltage configuration (TT, \SI{0.65}{\volt}, \SI{25}{\celsius}).
The extended design includes a partial sum buffer, an activation unit, and the \gls{hwpe} components.
The complete cluster requires \SI{0.991}{\square\milli\meter} (5 MGE) with the \gls{hwpe} subsystem occupying \SI{39.3}{\percent} of the total area.
The longest paths of the design are located between the input to the output of the dot product units in the \gls{hwpe}, within the \gls{dma}, and the instruction cache to the data mover core with gate delays of 12, 11, and 11, respectively.

\subsection{Neural Network Deployment}\label{sec:implementation:deployment}
To extend Deeploy with our architecture template, including the cluster and \gls{ita}, the mapping process of \gls{ita}-compatible operators is implemented in a multi-step approach. 
Deeploy starts by matching an \gls{mha} pattern and fuses it to form a monolithic node in the graph.
This node is then split along the head dimension to map the \gls{mha} operator head-by-head on \gls{ita}. Finally, a head accumulation layer is inserted at the end, which runs on the cluster cores.

As described in Section~\ref{sec:deploymentframework}, we extend Deeploy with a model for \gls{ita} to support HW-specific optimizations.
To solve the tiling problem, we specify geometrical tiling constraints to ensure all inputs and outputs have shapes compatible with \gls{ita}'s requirements. 
In the kernel, we preprogram the next tile using the dual-context register file and configure \gls{ita} to load the weights for the next step in the current one.
This enables us to achieve a fully double-buffered dataflow without starvation.

To the best of our knowledge, this is the first deployment flow that supports the \gls{e2e} acceleration of Attention-based \transformers{} at the edge.

%% file: src/50_Results.tex
\section{Results}
\label{sec:result}

\begin{table*}[ht]
\centering
\begin{threeparttable}
\caption{End-To-End Network Performance Metrics and Comparison to DNNs on Commercial tinyML Devices}
\label{tab:merged_result}
\input{tab/tab_e2e_merged}

\end{threeparttable}
\end{table*}

To measure the power consumption and latency of deployed workloads on our design, we perform cycle-accurate post-layout simulation of the entire cluster using Siemens QuestaSim for latency and throughput evaluation at \SI{425}{\mega\hertz} and post-layout gate-level simulations for power measurement under typical conditions (TT, \SI{0.65}{\volt}, \SI{25}{\celsius}). We choose the \SI{0.65}{\volt} operating corner to maximize energy efficiency. Our simulation setup accounts for latency and energy costs of memory transfers between the L1 and the system's background memory via the \gls{dma}, programming of the accelerator and cores, and execution of the operators, both on the cluster and \gls{ita}. 
In the following sections, we profile representative microbenchmarks and the execution of three different \transformer{} networks. Finally, we compare our results with state-of-the-art \gls{mcu}-class heterogeneous \glspl{soc} for \gls{tinyML}.

\subsection{Microbenchmarking Result}\label{sec:result_microbenchmark}
We analyze the performance and efficiency of \gls{gemm} and the more complex Attention operations and compare the multi-core cluster without any accelerator with the \gls{ita} integrated cluster.
Our heterogenous cluster achieves a throughput of \SI{741}{\giga\op\per\second} and energy efficiency of \SI{5.42}{\tera\op\per\joule} in \gls{gemm} computation, corresponding to $986 \times$ and $188 \times$ improvement respectively compared to the cluster without \gls{ita} with a peak accelerator utilization of \SI{85.1}{\percent}.
Running single-head Attention operation offers an even higher performance improvement of more than 3 orders of magnitudes and a $901 \times$ better energy efficiency resulting in \SI{663}{\giga\op\per\second} and \SI{6.35}{\tera\op\per\joule} with \SI{74.9}{\percent} accelerator utilization.
The standalone accelerator achieves a slightly higher utilization of \SI{79.6}{\percent}, with the integration into the template incurring only a small decrease of \SI{4.7}{p{.}p{.}}. This demonstrates that the template has minimal impact on the accelerator utilization.
This trend can be attributed to the efficient \softmax{} implementation in \gls{ita}, which does not add latency and thus avoids bottlenecking the overall efficiency.

\subsection{End-To-End Deployment Results}
To benchmark the execution of a complete model, we quantize MobileBERT\footnote{$S=128$, $E=128$, $P=64$, $H=4$, $N=24$, $d_{ff}=512$ (Sequence Length, Embedding Size, Projection Dimension, Attention Heads, Layers, Feed-Forward)}, DINOv2\footnote{$S=241$, $E=384$, $P=64$, $H=6$, $N=12$, $d_{ff}=1536$} and Whisper's\footnote{$S=512$, $E=384$, $P=64$, $H=6$, $N=4$, $d_{ff}=1536$} encoder using the QuantLib\footnote{\url{https://github.com/pulp-platform/quantlib}} library to perform 8-bit full integer inference.
Due to the extensive simulation time, we measure each layer separately and sum their execution times to extrapolate to the entire network. 
\autoref{tab:merged_result} display the \gls{e2e} results for two scenarios: multi-core cluster without the accelerator and multi-core cluster with the \gls{ita} accelerator.

In the scenario with a multi-core cluster, using \gls{ita} improves throughput up to $208 \times$ at $102 \times$ higher energy efficiency.

\subsection{Comparison with the State-of-the-art}\label{sec:result_soa}
To compare our work with the state-of-the-art in \gls{tinyML} computer architectures, we present the throughput and energy efficiency for various devices
in~\autoref{tab:merged_result}.
Due to the lack of \gls{e2e} benchmarks for \transformers{} on similar devices, we compare against \glspl{cnn} instead.
It is important to note that \transformers{} pose a greater challenge for accelerators due to their complex dataflow and computational demands. 

The Syntiant NDP120\footnote{\url{https://www.syntiant.com/hardware}} \gls{mcu} implemented in UMC \SI{40}{\nano\meter} ULP technology uses the Syntiant Core 2 tensor processor coupled with an Arm Cortex M0 processor and a HiFi-3 \gls{dsp}.
It achieves up to \SI{7}{\giga\op\per\second} at \SI{400}{\giga\op\per\joule} in \textit{MLPerf Tiny Inference} on MobileNetV1~\cite{banbury_mlperf_2021}.
We also compare with the Ensemble E3 \gls{ai} \gls{mcu} from Alif Semiconductor\footnote{\href{https://alifsemi.com/faster-ai-mcu-inferencing-low-power-consumption/}{https://alifsemi.com/}} which couples Ethos-U55 \gls{ml} processors with ARM Cortex M55 processors.
Depending on the network it achieves up to \SI{45}{\giga\op\per\second} at \SI{560}{\giga\op\per\joule}.
Compared to both devices, we achieve at least $3.4 \times$ more throughput with a $5.3 \times$ higher energy efficiency.

A comparison with a very similar architecture is possible against GreenWaves GAP9 \gls{mcu} containing the NE16 neural engine.
The \gls{soc} implemented in \SI{22}{\nano\meter} technology contains a fabric controller and a compute cluster with nine RISC-V cores and \SI{128}{\kilo\byte} shared L1 memory.
In the \textit{MLPerf Tiny Inference} benchmark on MobileNetV1 it achieves \SI{25}{\giga\op\per\second} at \SI{480}{\giga\op\per\joule} while Moosmann et al.~\cite{moosmann_ultra-efficient_2023} report better numbers with up to \SI{60}{\giga\op\per\second} at \SI{650}{\giga\op\per\joule} for a different network.
In comparison, we achieve $2.6 \times$ more throughput and $4.6 \times$ higher energy efficiency even though we deploy a more complex network.

%% file: tab/tab_e2e_merged.tex
\begin{tabularx}{\linewidth}{@{}lXcccccc@{}}
    \toprule
     & & \multicolumn{2}{c}{\textbf{Ours}} & \multicolumn{3}{c}{\textbf{Commercial Devices}} \\
    \cmidrule(lr){3-4} \cmidrule(lr){5-7}
    \textbf{Metric} & \textbf{Unit} & \textbf{Multi-Core}  & \textbf{Multi-Core + ITA} & \textbf{Syntiant NDP120}\tnote{$\,$\textdaggerdbl}~\cite{banbury_mlperf_2021} & \textbf{AlifSemi E3}\tnote{$\,$\S} & \textbf{GreenWaves GAP9}\tnote{$\,$*$\,$\textdaggerdbl}~\cite{moosmann_ultra-efficient_2023,banbury_mlperf_2021}  \\
    \midrule
    \textbf{Throughput} & [GOp/s]        & 0.74      & \textbf{56-154}                  & 2-7                   & 2-45                  & 10-60                \\
    \textbf{Energy Efficiency} & [GOp/J] & 28.9       & \textbf{1600-2960}                 & 280-400               & 50-560                & 150-650              \\
    \textbf{Power} & [mW]               & \textbf{26.0}       & 35.2-52.0                 & -                     & -                     & -                    \\
    \midrule
\end{tabularx}
\begin{tabularx}{\linewidth}{@{}lXcccccc@{}}
     & & \multicolumn{2}{c}{\textbf{MobileBERT}\tnote{$\,$a}} & \multicolumn{2}{c}{\textbf{DINOv2-Small}\tnote{$\,$b}} & \multicolumn{2}{c}{\textbf{Whisper-Tiny Encoder}\tnote{$\,$c}} \\
    \cmidrule(lr){3-4} \cmidrule(lr){5-6} \cmidrule(lr){7-8}
    \textbf{Metric} & \textbf{Unit} & \textbf{Multi-Core}  & \textbf{Multi-Core + ITA} & \textbf{Multi-Core}  & \textbf{Multi-Core + ITA} & \textbf{Multi-Core}  & \textbf{Multi-Core + ITA} \\
    \midrule
    \textbf{Energy per Inference} & [mJ/Inf]     & 164       & 1.60                 & 407                     & 7.31                     & 340            & 5.55           \\
    \textbf{Inference per Second} & [Inf/s]    &  0.16      & 32.5                 & 0.06                     & 4.83                     & 0.08          & 6.52             \\
    \bottomrule
\end{tabularx}
\begin{tablenotes}
    \item[\textdaggerdbl] MobileNetV1(x0.25) with \SI{28}{\mega\op}
    \item[\S] MicroNet Medium, MobileNetV2 1.0, Yolo-Fastest v4, Tiny Wav2letter Pruned, \href{https://alifsemi.com/faster-ai-mcu-inferencing-low-power-consumption/}{https://alifsemi.com/}
    \item[*] TinyissimoYOLO
    \item[a] \SI{4.74}{\giga\op} per inference with sequence length $S=128$
    \item[b] \SI{11.7}{\giga\op} per inference with sequence length $S=241$
    \item[c] \SI{9.74}{\giga\op} per inference with sequence length $S=512$
\end{tablenotes}

%% file: src/60_Conclusion.tex
\section{Conclusion}
\label{sec:conclusion}

We presented a flexible hardware-software architecture template, enabling collaborative accelerated execution of emerging Attention-based workloads that can be easily extended for the demands of future networks. 
By integrating our hardware template in Deeploy, we demonstrate a flexible deployment flow capable of efficiently mapping both accelerator-specific and generic \gls{dnn} operators on our target architecture.
We demonstrate the first \gls{e2e} deployment of multiple \transformer{}-based encoder models on a parallel heterogeneous accelerator-enhanced \gls{mcu}.
Our implementation, which leverages \gls{ita} for computing the \gls{mha} and Linear layers and the cluster cores for auxiliary operators, achieves state-of-the-art throughput of \SI{154}{\giga\op\per\second} with an energy efficiency of \SI{2.96}{\tera\op\per\joule}.
This enables inference rates of \SI{32.5}{\inference \per \second} at \SI{1.60}{\milli\joule \per \inference} for MobileBERT, \SI{4.83}{\inference \per \second} at \SI{7.31}{\milli\joule \per \inference} for DINOv2-Small, and \SI{6.52}{\inference \per \second} at \SI{5.55}{\milli\joule \per \inference} for encoder block of Whisper.

%% file: src/90_Acknowledgement.tex
\section*{Acknowledgment}
We thank Andrei Deaconeasa, Maximilian Coco, and Timon Fercho for their valuable contributions to the research project.
This work is supported in part by CONVOLVE (g.a. 101070374) and NeuroSoC (g.a. 101070634) projects under the European Union’s Horizon research and innovation programme, and TRISTAN (g.a. 101095947) project funded by Chips JU.

%% file: src/95_Authors.tex
\newpage
\begin{IEEEbiographynophoto}{Philip Wiese} received the B.Sc. and M.Sc. degree in electrical engineering and information technology from ETH Zürich in 2021 and 2023, respectively, where he is currently pursuing a Ph.D. degree at the Integrated Systems Laboratory. His research interests include machine-learning compilers and digital low-power design.
Contact him at wiesep@iis.ee.ethz.ch.
\end{IEEEbiographynophoto}
\vspace{\skipAuthor}
\begin{IEEEbiographynophoto}{Gamze \.{I}slamo\u{g}lu} received her B.Sc. degrees in Electrical and Electronics Engineering, and Physics from Bo\u{g}azi\c{c}i University in 2020, and M.Sc. degree in Electrical Engineering and Information Technology from ETH Zürich in 2022. She is currently pursuing a Ph.D degree at the Integrated Systems Laboratory at ETH Zürich. Her research interests include hardware accelerators for machine learning and heterogeneous multicore SoCs. Contact her at gislamoglu@iis.ee.ethz.ch.
\end{IEEEbiographynophoto}
\vspace{\skipAuthor}
\begin{IEEEbiographynophoto}{Moritz Scherer} received the B.Sc. and M.Sc. degree in electrical engineering and information technology from ETH Zürich in 2018 and 2020, respectively, where he is currently pursuing a Ph.D. degree at the Integrated Systems Laboratory. His research interests include ultra-low power and energy-efficient circuits and embedded design for machine learning. Contact him at scheremo@iis.ee.ethz.ch.
\end{IEEEbiographynophoto}
\vspace{\skipAuthor}
\begin{IEEEbiographynophoto}{Luka Macan} received the B.Sc. and M.Sc. degree from the Faculty of Electrical Engineering and Computing, University of Zagreb, Croatia in 2017 and 2019, respectively. He is currently pursuing a Ph.D. degree at the University of Bologna. His research interests include machine learning on embedded systems and hardware accelerators. Contact him at luka.macan@unibo.it.
\end{IEEEbiographynophoto}
\vspace{\skipAuthor}
\begin{IEEEbiographynophoto}{Victor Jean-Baptiste Jung}
 received his Bachelor's degree in Computer Science and Engineering Physics from Juniata College, and the Master's degree in Computer Science from the Institut Supérieur de l’Electronique et du Numérique of Lille (ISEN Lille) in 2022. He is currently pursuing a Ph.D. degree at the Integrated Systems Laboratory at ETH Zurich. His current research interests include the efficient deployment of ML models on microcontrollers and quantization. Contact him at jungvi@iis.ee.ethz.ch.
\end{IEEEbiographynophoto}
\vspace{\skipAuthor}
\begin{IEEEbiographynophoto}{Alessio Burrello}
received his M.Sc. and Ph.D. degrees in Electronic Engineering at the Politecnico of Turin, Italy, and the University of Bologna, respectively, in 2018 and 2023. He is currently working as a research assistant at Politecnico di Torino. His research interests include parallel programming models for embedded systems and hardware-oriented deep learning. Contact him at alessio.burrello@polito.it.
\end{IEEEbiographynophoto}
\vspace{\skipAuthor}
\begin{IEEEbiographynophoto}{Francesco Conti} received the Ph.D. degree in electronic engineering from the University of Bologna, Italy, in 2016. He is currently a Tenure-Track Assistant Professor with the DEI Department, University of Bologna. His research interests include hardware acceleration in ultra-low power SoCs for artificial intelligence applications. Contact him at f.conti@unibo.it.
\end{IEEEbiographynophoto}
\vspace{\skipAuthor}
\begin{IEEEbiographynophoto}{Luca Benini} holds the Chair of Digital Circuits and Systems at ETH Zürich and is a Full Professor with the Universit\`{a} di Bologna. He is a Fellow of the ACM and the IEEE and a member of the Academia Europaea. His research interests include energy-efficient computing systems and machine-learning hardware.
Contact him at lbenini@iis.ee.ethz.ch.
\end{IEEEbiographynophoto}